# A Potential Cathode Material for Rechargeable Potassium-Ion Batteries Inducing Manganese Cation and Oxygen Anion Redox Chemistry: Potassium-Deficient $K_{0.4}Fe_{0.5}Mn_{0.5}O_2$

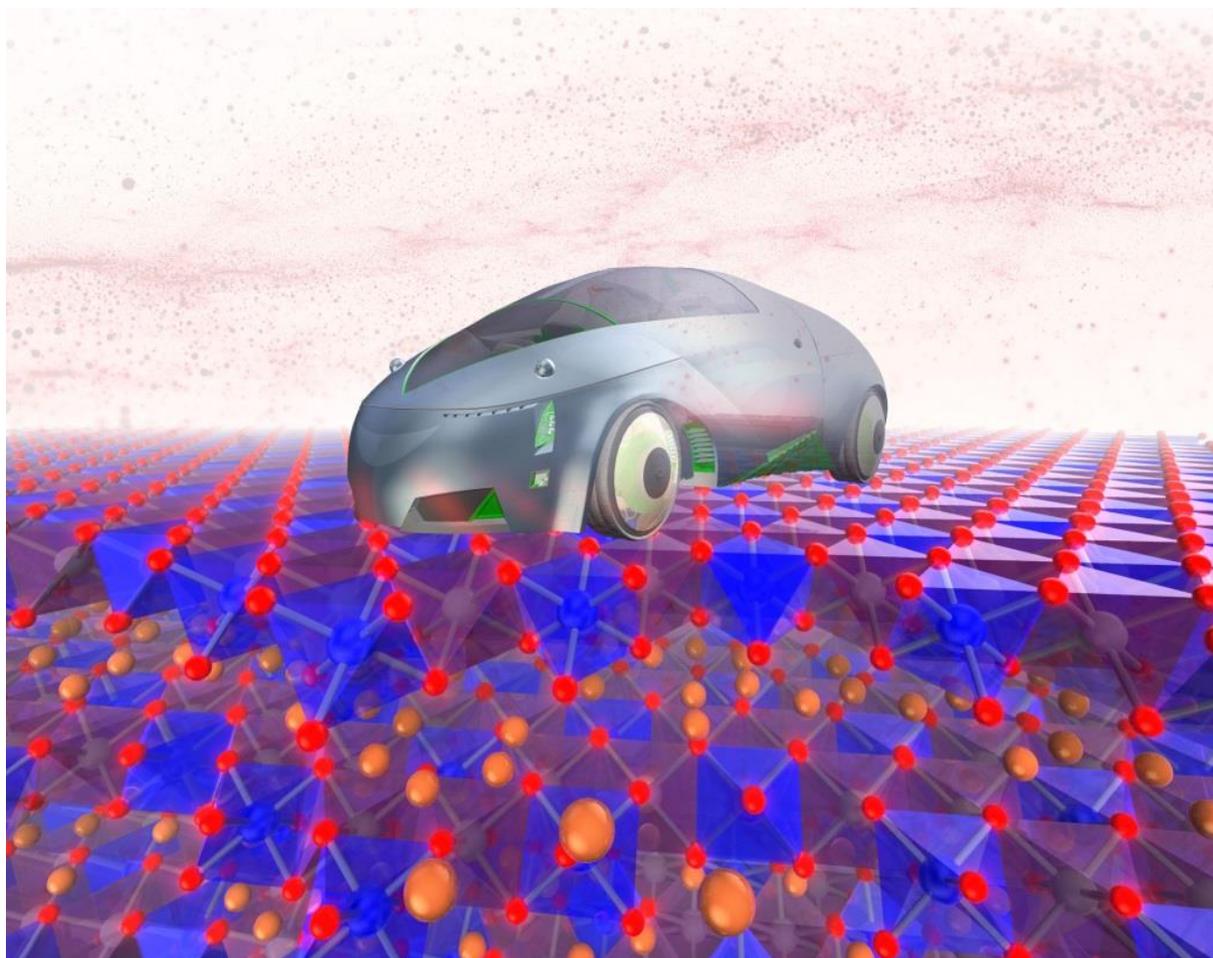

**Let Oxygen Also Flow!** A layered oxide cathode has been identified in the $K_2O$–$Fe_2O_3$–$MnO_2$ ternary phase system that shows facile and reversible potassium-ion (K-ion) mobility via a cumulative participation of cationic and anionic redox reaction, paving the way for the further development of high-energy-density cathode materials for K-ion batteries that effectively rely on both cation and anion multi-redox chemistry.

*Titus Masese[a]\*, Kazuki Yoshii[a], Kohei Tada[a], Minami Kato[a], Satoshi Uchida[a], Keigo Kubota[b], Toshiaki Ina[c], Toyoki Okumura[a], Zhen-Dong Huang[d]\*, Junya Furutani[e], Yuki Orikasa[e]\*, Hiroshi Senoh[a], Shingo Tanaka[a] and Masahiro Shikano[a]*

Dr. T. Masese, Dr. K. Yoshii, Dr. K. Tada, Dr. M. Kato, Dr. S. Uchida, Dr. T. Okumura, Dr. H. Senoh, Dr. S. Tanaka, Dr. M. Shikano
[a] Research Institute of Electrochemical Energy, Department of Energy and Environment (RIECEN), National Institute of Advanced Industrial Science and Technology (AIST), 1–8–31 Midorigaoka, Ikeda, Osaka 563–8577, JAPAN.
E-mail: titus.masese@aist.go.jp (Lead contact)

Dr. K. Kubota
[b] AIST-Kyoto University Chemical Energy Materials Open Innovation Laboratory (ChEM-OIL), Sakyo–ku, Kyoto 606–8501, JAPAN.

Dr. T. Ina
[c] Japan Synchrotron Radiation Research Institute, 1–1–1 Kouto, Sayo–cho, Sayo–gun, Hyogo 679–5198, JAPAN.

Prof. Z. –D. Huang
[d] Key Laboratory for Organic Electronics and Information Displays & Jiangsu Key Laboratory for Biosensors, Institute of Advanced Materials (IAM), Jiangsu National Synergetic Innovation Center for Advanced Materials (SICAM), Nanjing University of Posts and Telecommunications, 9 Wenyuan Road, Nanjing 210023, P.R. CHINA.

Mr. J. Furutani, Prof. Y. Orikasa
[e] Department of Applied Chemistry, College of Life Sciences, Ritsumeikan University, 1–1–1 Noji-higashi, Kusatsu, Shiga 535–8577, JAPAN.

**Abstract:**

Potassium-ion (K-ion) rechargeable batteries; considered to be lucrative low-cost battery options for large-scale and capacious energy storage systems, have been garnering tremendous attention in recent years. However, due to the scarcity of cathode materials that can condone the reversible re-insertion of the large K-ions at feasible capacities, the viability of K-ion batteries has been greatly undercut. In this paper, we explore a potential cathode material in the $K_2O$–$Fe_2O_3$–$MnO_2$ ternary phase system, that not only demonstrates reversible K-ion reinsertion but also manifests relatively fast rate capabilities. The titled cathode compound, $K_{0.4}Fe_{0.5}Mn_{0.5}O_2$, demonstrates a reversible capacity of approximately 120 mAh g$^{-1}$ at 10 hours of (dis)charge (*viz.*, C/10 rate) with *ca*. 85% of the capacity being retained at a 1 hour of (dis)charge (1 C rate) which is considered to be good capacity retention. Additionally, both hard and soft X-rays have been employed to unravel the mechanism by which K-ion is reversibly inserted into $K_{0.4}Fe_{0.5}Mn_{0.5}O_2$. The results revealed a cumulative participation of both manganese cations and oxygen anions in $K_{0.4}Fe_{0.5}Mn_{0.5}O_2$ illustrating its potential as a high-capacity K-ion battery cathode material that relies on both anion and cation redox. Further development of related high-capacity cathode compositions can be anticipated.

# 1. Introduction

Currently research on low-cost energy storage technologies has become momentous as energy and materials scientists relentlessly pursue the actualization of a sustainable energy society. Lithium-ion (Li-ion) battery technology has dominated the energy storage sector, with adoption ranging from portable electronics (*e.g*., mobile electronic gadgets such as smartphones, bioelectronics devices such as artificial cochlear implants, *etc*.), extending to large-scale energy storage systems such as the electric grid coupled with intermittent renewable energy such as solar and wind. However, geopolitical concerns over the supply of lithium resources as well as the imminent scarcity of the terrestrial reserves of lithium, undermine the sustainability of rechargeable Li-ion battery technology as the *numero uno* energy storage system. [1]

In this context, rechargeable batteries that utilize earth-abundant materials such as potassium-ion (K-ion) are gaining tremendous attention as not only promising low-cost alternatives to lithium-ion technology, but also as high-performance energy storage systems. [2] An economical and well-studied material such as graphite can be used directly as anode, while high voltages can be handled by coupling graphite with suitable cathode candidates and stable potassium-based electrolytes. [3-13] Despite the alluring virtues of adopting the K-ion technology, only a few electrode materials that can facilitate reversible insertion of the large potassium-ion exist. While extensive research on anode materials has yielded a plethora of potential anodes such as graphitic carbon materials, [3, 12] identifying cathode materials that can facilitate reversible K-ion re-insertion remains a formidable challenge. [4]

Layered oxides that mimic stoichiometric compositions akin to cathode materials of classic lithium- and sodium-ion batteries (such as $K_{0.6}CoO_2$, $KCrO_2$, $K_{0.3}CoO_2$ and so forth) have thus been brought to the forefront as promising cathode candidates. [6, 10] This comes as a result of their expected performance similarities with commercial Li-ion battery cathodes and their high theoretical volumetric energy densities. For instance, reversible K-ion re-insertion at moderate capacities in conjunction with fast rate capabilities was observed in layered metal oxides incorporating mixed-transition metals such as $K_{0.7}Fe_{0.5}Mn_{0.5}O_2$, $K_{0.67}Ni_{0.16}Co_{0.16}Mn_{0.67}O_2$ and $K_{0.65}Fe_{0.5}Mn_{0.5}O_2$.[14-16] Indeed they show enhanced electrochemical performance compared to the parent $K_xMnO_2$ phases. This has sparked renewed interest in other new layered cathode materials such as iron- and manganese-based layered cathode frameworks with greater potentials.

A vast majority of the research focusing on iron-based cathode materials for rechargeable cation batteries tends to favor materials that consist of trivalent iron ($Fe^{3+}$); therefore $KFeO_2$ would be a suitable candidate, as the $Fe^{3+}/Fe^{4+}$ redox couples are expected to be utilized to attain high capacity, or presumably illustrate the cumulative participation of both iron cations and oxygen anions to the redox processes as has been exemplified in $NaFeO_2$.[17] Nevertheless, $KFeO_2$ adopts the cristobalite framework with $Fe^{3+}$ in tetrahedral coordination (see **Figure S1** in the **Supplementary Information** section) and drastically undergoes degradation upon the initial K-ion extraction (charging) process (see **Figure S2**). To stabilize $Fe^{3+}$ in the $KFeO_2$ framework during the oxidation process, it would be befitting to utilize congener transition metal ions that have essentially the same Shannon-Prewitt ionic radii as $Fe^{3+}$ and that can more covalently bond with oxygen to avert oxygen loss or release during oxidation.[17, 18] On this account, we sought to identify new mixed-transition metal cathode oxides in the $K_2O–Fe_2O_3–MnO_2$ ternary phase system (shown in **Figure S3**).

Herein, we report on a new potassium-deficient $K_{0.4}Fe_{0.5}Mn_{0.5}O_2$ cathode oxide (of which '$K_{0.5}Fe_{0.5}Mn_{0.5}O_2$' was the targeted phase composition) comprising predominantly of trivalent iron ($Fe^{3+}$) and tetravalent manganese ($Mn^{4+}$). This in stark contrast to the Fe– and Mn–based cathode oxides reported so far that comprise both trivalent iron ($Fe^{3+}$) and manganese ($Mn^{3+}$).[15, 16] $K_{0.4}Fe_{0.5}Mn_{0.5}O_2$ possesses a layered structure and exhibits a reversible specific capacity of ~120 mAh g$^{-1}$ as well as good capacity retention. Hard and soft X-ray measurements reveal the cumulative participation of transition metal cations and oxygen anions during the K-ion extraction and reinsertion process, paving the way for the advancement of mixed-transition metal oxide cathode materials for K-ion batteries that rely on anion hole-cation redox chemistry to attain high capacity.

## 2. Results and Discussion

$K_{0.4}Fe_{0.5}Mn_{0.5}O_2$ was synthesized using a scalable solid-state reaction of $K_2CO_3$, $Fe_2O_3$ and $MnO_2$ as the starting precursors. Further details are furnished in the **Experimental** section. **Figure 1a** shows a high-quality synchrotron X-ray diffraction XRD (SXRD) pattern of the as-prepared $K_{0.4}Fe_{0.5}Mn_{0.5}O_2$, tentatively indexed in an orthorhombic unit cell in the space group *Ccmm* with the following cell parameters: $a = 5.2434(5)$ Å, $b = 2.8821(2)$ Å, $c = 13.8482(1)$ Å ($V = 209.28(1)$ Å$^3$). Additional details relating to unit cell parameters and atomic coordinates are provided in the **Supplementary Information** section (**Table S1**). Transmission electron microscopy (TEM) and scanning electron microscopy (SEM) were employed to collect details regarding the morphological aspects of the as-prepared $K_{0.4}Fe_{0.5}Mn_{0.5}O_2$. TEM images, shown in **Figures 1b** and **1c**, validate the orthorhombic symmetry with clear diffraction spots arising from the high crystallinity of the obtained powder. This is in good accord with the simulated diffraction patterns shown in **Figure 1d**. Further, powder morphology of the obtained material is provided in **Figure 1e**. Despite the accretion of a few small crystallites, a uniform distribution of particles is observed in the sample. The elemental analyses (shown in **Figures 1f and S4**) indicate the presence of potassium (K), iron (Fe) and manganese (Mn). Inductively coupled plasma (ICP) measurements (see **Table S2**) reveal an atomic distribution of K, Fe and Mn in the ratio 0.39: 0.5:0.49, respectively, translating to a K-deficient composition of $K_{0.4}Fe_{0.5}Mn_{0.5}O_2$. Hard X-rays, particularly X-ray absorption near-edge structures (XANES) spectra given in **Figure S5**, further confirm Fe and Mn to be predominantly in the trivalent ($Fe^{3+}$) and tetravalent ($Mn^{4+}$) state, as should be expected.

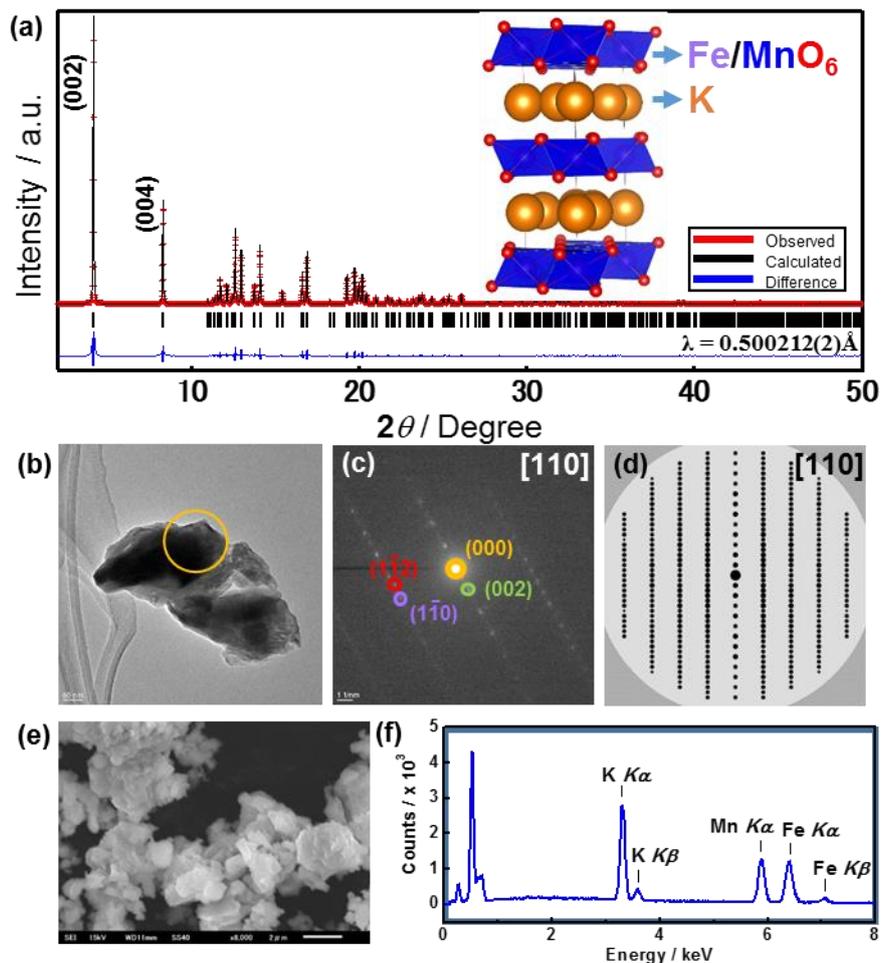

**Figure 1.** Physicochemical characterization of layered $K_{0.4}Fe_{0.5}Mn_{0.5}O_2$. **(a)** Rietveld refinement of high-quality synchrotron XRD pattern of the pristine $K_{0.4}Fe_{0.5}Mn_{0.5}O_2$ ($R_B$ = 3.71%, $\chi^2$ = 1.25). The observed and calculated peaks are indicated in red and black, respectively. The difference between the observed and calculated intensity is indicated in blue, while black ticks indicate the position of the Bragg peaks of the P-type layered phase indexed in the *Ccmm* orthorhombic space group. Here P-type coordination denotes prismatic coordination of the K-ions in brucite-like layers of the unit cell. **(b)** Transmission electron microscopy (TEM) images of $K_{0.4}Fe_{0.5}Mn_{0.5}O_2$ and **(c)** the corresponding selected area electron diffraction (SAED) patterns indexed in an orthorhombic unit cell along [110] zone axis. **(d)** SAED simulations affirming the indexing of the experimental diffraction patterns. **(e)** SEM images of micrometric particles and **(f)** energy dispersive X-ray (EDX) spectra confirming the elemental composition of the as-prepared $K_{0.4}Fe_{0.5}Mn_{0.5}O_2$. A more succinct spectra is furnished in the Supplementary Information section (Figure S4).

To assess the electrochemical properties of the as-prepared $K_{0.4}Fe_{0.5}Mn_{0.5}O_2$, galvanostatic (dis)charge measurements and cyclic voltammetry were conducted in potassium half-cells.

Main redox voltages, centered at around 2.8 V, are observed in the cyclic voltammograms (shown in **Figure 2a**). Note that an ionic liquid based on potassium bis(trifluoromethanesulfonyl)amide (KTFSA) salt was used as the electrolyte, as it previously demonstrated stability during high-voltage regime operations.[13] **Figure 2b** shows the voltage (dis)charge curves of $K_{0.4}Fe_{0.5}Mn_{0.5}O_2$ at room temperature. In spite of the high viscosity innate in KTFSA-based ionic liquids, $K_{0.4}Fe_{0.5}Mn_{0.5}O_2$ sustains a relatively high capacity of ~120 mAh $g^{-1}$ at a current density commensurate to C/20 rate (that is, 20 h of (dis)charge to full capacity; in this case *ca*. 6 mA $g^{-1}$). It would be important to mention that higher capacity is attained during initial discharge rather than during the initial charge, which is not unprecedented as the initial charge process starts from the potassium-deficient $K_{0.4}Fe_{0.5}Mn_{0.5}O_2$ phase. Moreover, the differential voltage-capacity plots (shown in **Figure S6**) are in good accord with those displayed in the cyclic voltammograms (**Figure 2a**). To further evaluate the rate performance of $K_{0.4}Fe_{0.5}Mn_{0.5}O_2$, galvanostatic (dis)charge measurements were performed using a potassium bis(fluorosulfonyl)amide (KFSA)-based electrolyte with a lower viscosity. [13, 19] Here, the upper cut-off voltage was limited to 3.8 V to avoid spurious reactions, further details of which are beyond the scope of the present study. **Figure 2c** shows the voltage-capacity plots at a C/10 rate, while the corresponding cyclability and rate capability data are shown in **Figure 2d**. $K_{0.4}Fe_{0.5}Mn_{0.5}O_2$ can sustain respectable rate capabilities and good capacity retention. A reversible capacity of approximately 120 mAh $g^{-1}$ at 10 hours of (dis)charge (*viz.*, C/10 rate) is still attainable (and at a coulombic efficiency of *ca.* 90% after 50 cycles), with approximately 85% of the capacity being retained after 1 hour of (dis)charge (1 C rate) showing a good capacity retention. Compared to some other reported Fe- and Mn-based layered cathode oxide materials (as succinctly summarized in **Figure 2e**),[4, 6, 14-16] $K_{0.4}Fe_{0.5}Mn_{0.5}O_2$ exhibits remarkably high-energy density and therefore a potential candidate for a low-cost and high-energy density K-ion battery cathode.

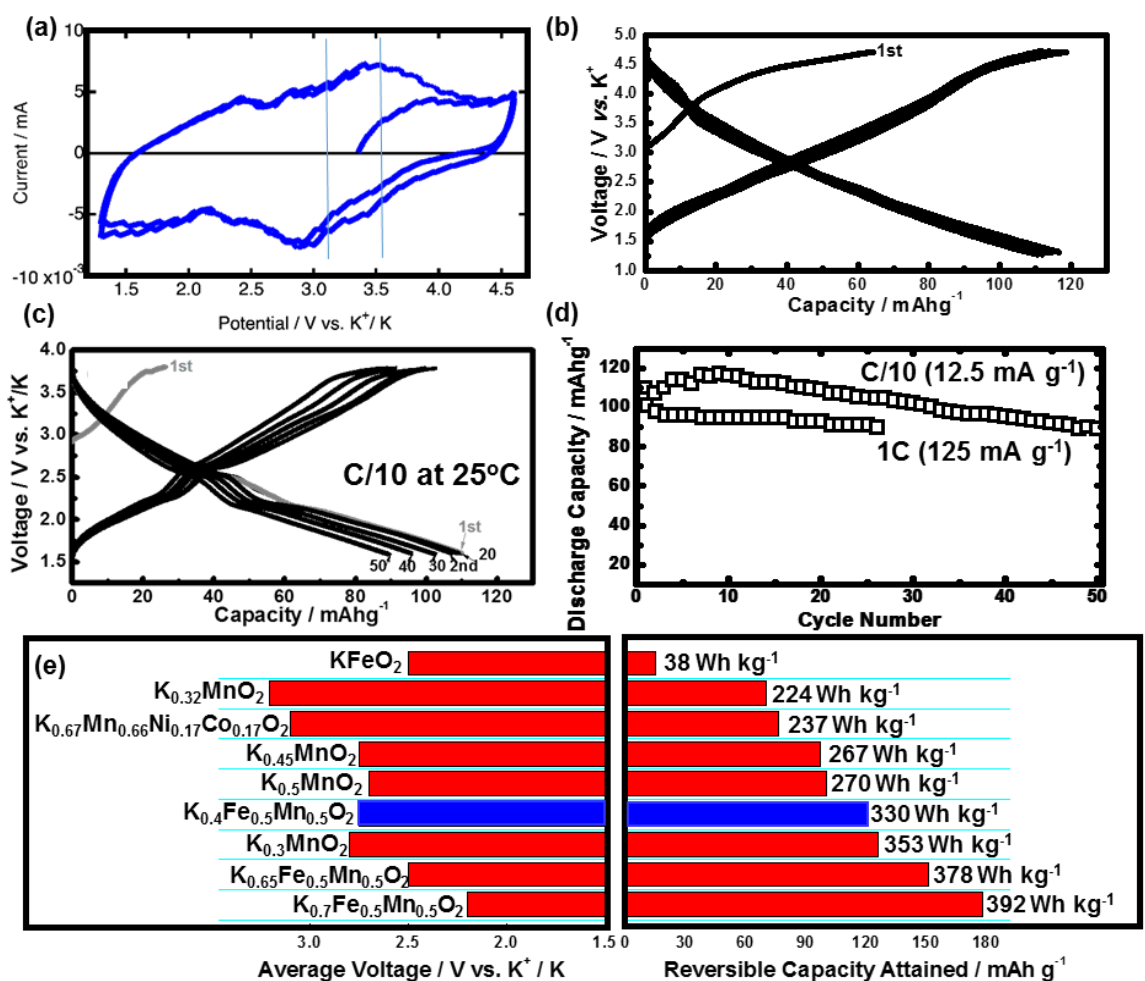

**Figure 2.** Electrochemical performance of $K_{0.4}Fe_{0.5}Mn_{0.5}O_2$. **(a)** Cyclic voltammograms of $K_{0.4}Fe_{0.5}Mn_{0.5}O_2$ at a scan rate of 0.1 mV s$^{-1}$ in K-half cells comprising 0.5 M KTFSA in Pyr$_{13}$TFSA ionic liquid as the electrolyte. **(b)** The corresponding voltage-capacity plots at a current density commensurate to C/20 (20 h of (dis)charge) and at a cut-off voltage of 4.7 V to 1.5 V. **(c)** Voltage-capacity plots in electrolyte consisting of 1 M KFSA in Pyr$_{13}$FSA ionic liquid (that affords lower viscosity and higher ionic conductivity than KTFSA-based ionic liquids) at a C/10 rate. Note that, as for the use of KFSA-based electrolyte, the upper cut-off voltage was set at 3.8 V to avert spurious reactions that were found to occur at higher cut-off voltages. **(d)** The cyclability data at varying current densities, demonstrating the relatively stable performance under respectable rate capabilities. **(e)** Bar plots benchmarking the performance of $K_{0.4}Fe_{0.5}Mn_{0.5}O_2$ with reported iron-and manganese-based oxide cathode materials for potassium-ion battery.

At this juncture, it is requisite to expound on transition metal cations participating in the charge compensation especially during the initial charging process, where an anomalous capacity is exhibited. Considering that no decomposition of the ionic liquid was observed within the set cut-off voltages, the anomalous capacity at the initial charging process can be envisioned to stem from the $Fe^{3+}/Fe^{4+}$ redox process alone since, at the onset of cycling, Mn initially exists in the tetravalent state ($Mn^{4+}$). To obtain unequivocal electronic information on the electrochemical redox process, *ex situ* X-ray absorption spectroscopy (XAS) measurements were collected at the Fe *K*– and Mn *K*–edges. Further details are provided in the **Experimental** section. Fe *K*–edge X-ray absorption near-edge structures (XANES) spectra during initial charging, initial discharging and subsequent charging processes are shown in **Figures 3a**, **3b** and **3c**. No conspicuous change in the absorption edge can be observed in both charging and discharging processes, indicating that the valence state of Fe (initially as $Fe^{3+}$, as evinced in **Figure S6**) essentially remains unaltered during the redox processes. Fe *K*–edge extended X-ray absorption fine structures (EXAFS) spectra also show no discernible changes (provided in **Figure S7** in the **Supplementary Information**), reinforcing the notion that Fe does not actively take part in the charge compensation process. Mn *K*–edge XANES spectra (shown in **Figures 3d, 3e** and **3f**) further reveal that, apart from the initial charging process where anomalous capacity is observed, the absorption edge shifts towards lower and higher energies, which is evident during initial discharging and second charging, respectively, due to the decrease and increase in the valence state of Mn cations. To determine the oxidation states of Mn during the redox process, comparison with Mn reference compounds (namely, $Mn_2O_3$ and $MnO_2$) was done (**Figure S8**); thus, verifying the reversible redox process of $Mn^{3+}$ and $Mn^{4+}$, respectively. Mn *K*–edge FT (Fourier Transform) EXAFS spectra additionally show a progressive decrease and increase in the average Mn–O and Mn–

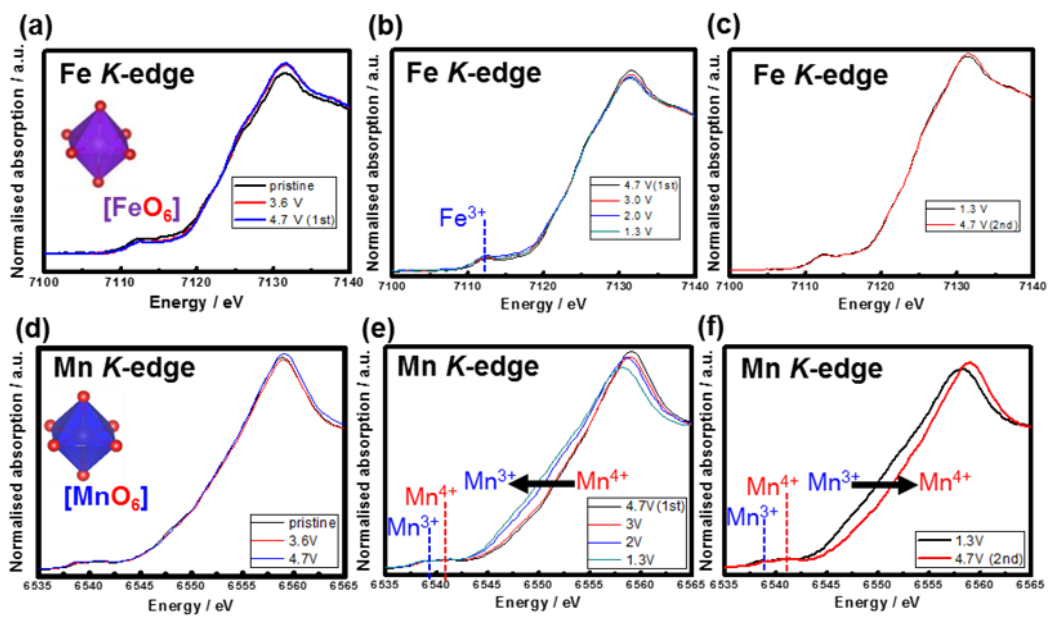

**Figure 3.** Analyses of the change in the valence state of iron (Fe) and manganese (Mn) cations in $K_{0.4}Fe_{0.5}Mn_{0.5}O_2$ during the initial charge and discharge processes (cycling) using hard X-rays. XANES spectra taken at Fe $K$–edge during **(a)** initial charging, **(b)** initial discharging and **(c)** second charging processes. XANES spectra at Mn K–edge during **(d)** initial charging, **(e)** initial discharging and (f) second charging processes. Distinct from Mn, the valence state of Fe remains virtually unaltered during both charging and discharging processes. Note that the valence of Fe is initially the trivalent state while Mn is at the tetravalent state, as confirmed further by pertinent reference compound (shown in Figures S6 and S8)

Mn bond lengths of the first neighboring shell during initial discharging and second charging (see **Figure S9**), respectively, in accord with the decrease and increase in the valence state of Mn cations. Noteworthy, is that the change in the valence state of Mn is more discernible at low-voltage regimes, implying that other charge compensation process should be occurring at high-voltage regimes and hence, the anomalous capacity. Based on the Fe $K$– and Mn $K$–edge XAS measurement results altogether, we conclude here that neither the oxidation of trivalent Fe ($Fe^{3+}$) nor tetravalent Mn ($Mn^{4+}$) occurs upon initial extraction of potassium (*viz.*, initial charging). This surmises that the oxidation of oxygen anions ($O^{2-}$ ions) also contributes to the charge compensation mechanism upon initial depotassiation (charging) of $K_{0.4}Fe_{0.5}Mn_{0.5}O_2$ and particularly at high-voltage regimes; hence the anomalous capacity is attainable.

To test this hypothesis, O $K$–edge XAS measurements were performed in order to collect information regarding the role of oxygen in the charge compensation mechanism, particularly at high-voltage regimes. XANES spectra at O $K$–edge of $K_{0.4}Fe_{0.5}Mn_{0.5}O_2$ during initial charging and discharging are shown in **Figures 4a** and **4b**, respectively. The data were collected using fluorescence yield (FY) mode, which can obtain unequivocal information particularly relating to the bulk. From the viewpoint of the localized nature of the oxygen 1$s$ orbital, the intense absorption peaks centered at 528–534 eV, in principle, represent the transition of the oxygen 1$s$ electron to the hole state in the oxygen 2$p$ orbital level hybridized with transition metal 3$d$ orbitals.[20] The broad and higher energy peaks above 536 eV are assignable to the transitions to hybridized states of oxygen 2$p$ and transition metal 4$sp$ orbitals.[20] Upon initial oxidation process (that is, charging), a peak emerges at around 531 eV while the absorption pre-edge spectra centered at around 529 eV shifts to lower energy regions. Even without a quantitative analysis, which is beyond the scope of this work, the apparent emergence of this peak upon K-ion extraction of $K_{0.4}Fe_{0.5}Mn_{0.5}O_2$ (attributed to the transition metal–3$d$ orbital states hybridized with O–2$p$ orbital states) strongly indicates that oxygen is redox-active at high-voltage regimes.[20] Indeed, well-known materials that contain $Fe^{4+}$, such as $CaFeO_3$, $LaCu_3Fe_4O_{12}$ and $SrFeO_3$, possess 3$d^5L$ *in lieu* of a 3$d^4$ electronic configuration.[21-23] For clarity, $L$ here represents an oxygen ligand hole, which means that oxygen is the active electron donor instead of iron. This observation is further affirmed by theoretical calculations of the potassium-deficient $K_{0.4-x}Fe_{0.5}Mn_{0.5}O_2$ phase that show the oxygen orbitals lying at the vicinity of the Fermi energy level (**Figure S10**). The reversibility of the anionic reaction of oxygen can be observed in the O $K$–edge spectra of $K_{0.4}Fe_{0.5}Mn_{0.5}O_2$ upon discharging. **Figure 4c** schematically shows the overall reaction mechanism that occurs during potassium-ion extraction and reinsertion in potassium-deficient $K_{0.4}Fe_{0.5}Mn_{0.5}O_2$. Potassium extraction process in $K_{0.4-x}Fe_{0.5}Mn_{0.5}O_2$ leads to, in principle, the oxidation of $Fe^{3+}$ ($t^3_{2g}$, $e^2_g$) to $Fe^{4+}$ ($t^3_{2g}$, $e^1_g$). However, since no further oxidation of $Fe^{3+}$ is observed experimentally, it can be assumed that such high-voltage regimes trigger the oxidation of oxygen ions such that $Fe^{4+}$ ions possess $3d^5L$ electronic structure with a ligand hole (denoted as $L$) being created in the oxygen anions. This is indeed validated by the O $K$–edge XAS measurement results, shown in **Figure 4a**. Moreover, upon potassium reinsertion, oxygen anions are reduced at high-voltage regimes while at low-voltage regimes $Mn^{4+}$ undergo reduction to $Mn^{3+}$ ions to yield a $K_{0.4+x}Fe_{0.5}Mn_{0.5}O_2$ composition.

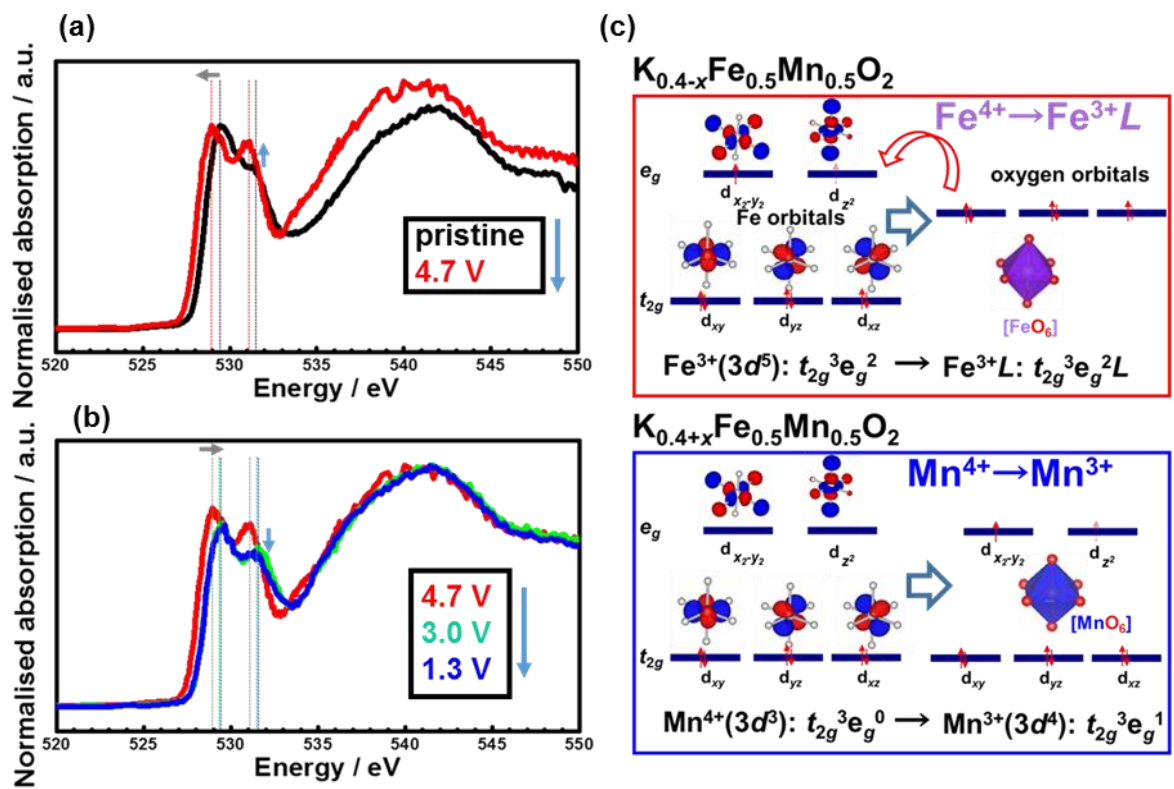

**Figure 4.** Analyses of the change in the valence state of oxygen anion in $K_{0.4}Fe_{0.5}Mn_{0.5}O_2$ during the initial charge and discharge processes (cycling) using soft X-rays. X-ray absorption near-edge structure (XANES) spectra at O $K$-edge taken during **(a)** initial charging and **(b)** initial discharging processes. The emergence of a new peak on the oxygen $K$-edge pre-edge features (centered around 531 eV) during $K^+$ extraction (charging) at high voltage regimes and its disappearance upon $K^+$ reinsertion, indicates the reversible charge compensation facilitated by oxygen during charging and discharging processes. **(c)** Illustration of the oxygen anion redox process occurring in $K_{0.4}Fe_{0.5}Mn_{0.5}O_2$ upon $K^+$ extraction at high-voltage regimes. Holes are created in the oxygen orbital.

At this point, a legitimate question pertains to the creation of holes in O–2$p$ orbitals; a risk is the evolution of oxygen which might be inimical to the structural integrity of $K_{0.4}Fe_{0.5}Mn_{0.5}O_2$ when used as an electrode. However, the excellent cyclability of $K_{0.4}Fe_{0.5}Mn_{0.5}O_2$ as well as the reversible crystal structural changes observed during initial charging and discharging processes (see **Figure S11** in the **Supplementary Information** section), prove otherwise. Nevertheless, without regard neither to the detailed chemistry nor physics underlying the hole-redox chemistry of oxygen (particularly whether electron transfer occurring from the 'oxidized' oxide ions involve the formation of the elusive oxo-like species such as super oxides ($O_2^-$) as reported in lithium-based compounds that display similar spectroscopic changes)[20], this study provides unambiguous experimental evidence that it is possible to achieve mixed-valence states (involving both constituent cations and anions) in potassium-deficient cathode materials for K-ion battery such as $K_{0.4}Fe_{0.5}Mn_{0.5}O_2$ exhibiting high capacity and excellent cyclability. Indeed, it is patent that the addition of $Mn^{4+}$ into $KFeO_2$ triggers the anion redox mechanism.

Additionally, this work has underpinned a new avenue for cathode materials based on the exploration of the $K_2O$–$Fe_2O_3$–$Mn^{4+}O_2$ ternary phase diagram where low-cost layered phases as $K_xFe_{0.5}Mn_{0.5}O_2$ ($x \leq 0.5$) can be anticipated to exist. We need to reiterate that this new cathode material we have identified comprises predominantly of trivalent iron ($Fe^{3+}$) and tetravalent manganese ($Mn^{4+}$), which distinguishes this work with that of Wang and co-workers (see **Table S3**).[15, 16] From a scalability point of view, $K_xFe_{0.5}Mn_{0.5}O_2$ can be easily prepared from a direct reaction of potassium iron oxide ($KFeO_2$) and manganese (IV) oxide ($MnO_2$), both of which happen to be earth-abundant materials (see **Figures S12** and **S13**). $KFeO_2$ is also a cheap salt used as a catalyst in polymer manufacturing or could be found as a by-product, formed by the reaction of (potassium oxide) $K_2O$ and iron (III) oxide ($Fe_2O_3$) in molten slags during refining of Fe. [24-25] On the other hand, $MnO_2$ is an abundant oxide comprising of oxide minerals examples of which are *pyrolusite*, *nsutite* (*yokosukalite*), *birnessite*, *hollandite*, *cryptomelane* (*ishiganeite*) and *ramsdellite*. [26] Therefore, preparation of $K_xFe_{0.5}Mn_{0.5}O_2$ is indubitably economical. There is an ongoing work pertaining to the assessment of the various $K_xFe_{0.5}Mn_{0.5}O_2$ ($x \leq 0.5$) polytypic phases so that a comprehensive $K_2O$–$Fe_2O_3$–$MnO_2$ ternary phase diagram is obtained.

## 3. Conclusion

In summary, a layered potassium-deficient cathode material in the $K_2O$–$Fe_2O_3$–$MnO_2$ ternary phase system has successfully been synthesized using a scalable solid-state method. This new cathode material, $K_{0.4}Fe_{0.5}Mn_{0.5}O_2$, can endure reversible potassium-ion (K-ion) reinsertion with a reversible capacity of *ca*. 120 mAh g$^{-1}$ at an average voltage of approximately 2.8 V. Further, $K_{0.4}Fe_{0.5}Mn_{0.5}O_2$ displays good capacity retention upon subsequent cycling with adequate rate capabilities. Hard and soft X-rays have been utilized to elucidate the complexity of the K-ion oxidation process in this material. The overarching results is the cumulative participation of not only the transition metal cations to the charge-compensation process, but also the hole-redox chemistry of oxygen to achieve a high specific capacity in $K_{0.4}Fe_{0.5}Mn_{0.5}O_2$–the first provision for a K-ion battery cathode material. We believe that this work will serve as a cornerstone for the development of new layered oxide materials for rechargeable potassium-ion battery that effectively rely on the cumulative cationic and anionic redox reaction to attain high capacity.

**Supporting Information**
Supporting Information is available via the following link:
https://onlinelibrary.wiley.com/doi/abs/10.1002/ente.202000039

**Acknowledgements**

We gratefully acknowledge Ms. Kumi Shiokawa and Ms. Yumi Haiduka for their advice and technical help as we conducted the electrochemical and XRD measurements. We thank Dr. Keiichi Osaka for conducting synchrotron X-ray diffraction measurements at SPring-8 facility (Proposal number 2017B1773).